\PassOptionsToPackage{table}{xcolor}
\documentclass[acmsmall, screen, nonacm]{acmart}
\usepackage[utf8]{inputenc}
\usepackage{xspace}
\usepackage{ifthen}
\usepackage[T1]{fontenc}
\usepackage{graphicx}
\usepackage{svg}
\usepackage{listings}
\usepackage{xcolor}
\usepackage{caption}
\usepackage{balance}
\usepackage{subcaption}
\usepackage{listings}
\usepackage{tabularx}
\usepackage{booktabs, multirow} 
\usepackage{algorithm}
\usepackage{algpseudocode}
\usepackage{numprint}
\usepackage[most]{tcolorbox}
\usepackage{url}
\usepackage{tikz}
\usepackage{makecell}

\usepackage[framemethod=tikz]{mdframed}
\mdfdefinestyle{mpdframe}{
    frametitlebackgroundcolor   =black!15,
    frametitlerule              =true,
    roundcorner                 =1pt,
    middlelinewidth             =1pt,
    skipabove                   =\baselineskip,
    innermargin                 =0.1cm,
    innerleftmargin             =0.1cm,
    innerrightmargin            =0.1cm,
    innertopmargin              =0.1cm,
    innerbottommargin           =0.1cm
}

\definecolor{color1bg}  {HTML}{9FC8C8}
\definecolor{color2bg}  {HTML}{D8A6A6}
\definecolor{color1line}{HTML}{1A80BB}

\title{Proving and Rewarding Client Diversity to Strengthen Resilience of Blockchain Networks}

\author{Javier Ron}
\email{javierro@kth.se}
\affiliation{%
  \institution{KTH Royal Institute of Technology}
  \city{Stockholm}
  \country{Sweden}
}

\author{Zheyuan He}
\email{zheyuanh@kth.se}
\affiliation{%
  \institution{KTH Royal Institute of Technology}
  \city{Stockholm}
  \country{Sweden}
}

\author{Martin Monperrus}
\email{monperrus@kth.se}
\affiliation{%
  \institution{KTH Royal Institute of Technology}
  \city{Stockholm}
  \country{Sweden}
}

\newcommand{\revised}[1]{{\color{black}#1}}

\begin{document}

\begin{abstract}
Client diversity is a cornerstone of blockchain resilience, yet most networks suffer from a dangerously skewed distribution of client implementations.
\revised{This monoculture exposes the network to very risky scenarios, such as massive financial losses in the event of a majority client failure.}

In this paper, we present a novel framework that combines verifiable execution and economic incentives to provably identify and reward the use of minority clients, thereby promoting a healthier, more robust ecosystem.
Our approach leverages state-of-the-art verifiable computation (zkVMs and TEEs) to generate cryptographic proofs of client execution, which are then verified on-chain.

We design and implement an end-to-end prototype of verifiable client diversity in the context of Ethereum, by modifying the popular Lighthouse client and by deploying our novel diversity-aware reward protocol. 
Through comprehensive experiments, we quantify the practicality of our approach, from overheads of proof production and verification to the effectiveness of the incentive mechanism.

This work demonstrates, for the first time, a practical and economically viable path to encourage and ensure provable client diversity in blockchain networks.
Our findings inform the design of future protocols that seek to maximize the resilience of decentralized systems.
\end{abstract}



\maketitle
\section{Introduction}
\label{sec_intro}
A blockchain is a distributed ledger in which information is saved in a sequence of blocks shared by all the operating nodes~\cite{nakamoto2008bitcoin}.
The participating nodes agree on the new blocks to be added to the blockchain by using a consensus algorithm. 
This entails a process where some of the nodes propose new blocks, while others validate and confirm them.
To incentivize participation, nodes are rewarded with some form of cryptocurrency~\cite{wood2014ethereum}.
For instance, in Ethereum, one of the largest blockchains with over 300 Billion USD market capitalization~\cite{eth_cap}, participating nodes receive \textit{Ether} when they participate to the blockchain protocol.

A prominent feature of blockchains are \textit{smart contracts}~\cite{smart-contracts}.
A smart contract is a computer program stored on the blockchain and whose execution outputs are all written to the blockchain as well.
The \textit{Decentralized Finance} (DeFi) ecosystem is built upon these smart contract-enabled blockchains~\cite{defi}, such as Ethereum.

Blockchains, as distributed systems with redundant data, are resilient to the failures of individual nodes. Yet, they are vulnerable to systemic software faults~\cite{klin_post3}
If the same bug is triggered across multiple nodes simultaneously, it could compromise the whole network, leading to significant reputation or financial losses~\cite{EthereumDocs}.
For instance, a bug~\cite{consensus_bug} in an Ethereum~\cite{wood2014ethereum} client was triggered in 2020, and led to the wrong acceptance of 30 blocks and involved a loss of around 8.6M USD~\cite{yang2021finding}.
In this paper, we aim at reducing the risks of systemic bugs or vulnerabilities. 

We look at software diversity as primary mechanism to mitigate systemic risks. Software diversity here means that there are multiple implementations of blockchain nodes, all talking the same protocol~\cite{monoculture}.
In a scenario with diverse blockchain node implementations, all nodes would never crash at the same time, because they do not share the same bugs and triggering conditions.
The Ethereum blockchain community is aware of this desirable property, and thus incentivizes so-called \textit{client diversity}~\cite{criticalneediversity}: having several, diverse, interoperable implementations of blockchain nodes, working together in the blockchain network.
Client diversity is real: the Ethereum consensus blockchain includes many diverse clients such as Prysm~\cite{prysm_client}, Lighthouse~\cite{lighthouse_client}, Teku~\cite{teku_client}, and Nimbus~\cite{nimbus_client}, providing resilience to Ethereum.
In this context, the following terminology is used by the community and in this paper:
\textit{minority clients} are run by less than $\frac{1}{3}$ of all nodes),  \textit{majority clients}  by more than $\frac{1}{3}$ and less than $\frac{2}{3}$ of all nodes, and \textit{super-majority clients} by more than $\frac{2}{3}$ of all nodes~\cite{klin_post3}.

Despite client diversity being considered crucial by the Ethereum community~\cite{criticalneediversity}, it is critically affected by a heavily skewed distribution of client implementations. Most nodes run the same code, i.e. there are majority clients, and blockchain stakeholders are terrified by the possibility of super-majority clients.
Assuming that a bug can be triggered in a super-majority client, the worst-case scenarios would happen: network partitions, storing incorrect data in the blockchain, financial loss, and day-long outages.
In this paper, we propose a novel solution to prevent catastrophes resulting from majority and super-majority clients.

Our key insight to systemic risks due to monoculture is a conceptual and technical framework that ensures client diversity beyond soft advocacy.
Our proposal solves the problem through a combination of verifiable execution~\cite{chen2022securing,ghodsi2017safetynets} and economic incentives.
The idea is to tune the uneven distribution of Ethereum clients through an economic incentive mechanism to use minority clients.
Our framework provides higher rewards to nodes that run \textit{minority clients} (the clients that have a smaller share).
To ensure the authenticity of minority clients claiming higher rewards, our framework employs verifiable execution. 
By leveraging verifiable execution techniques, any participant can independently verify that a node has indeed executed a particular client implementation, without requiring trust in the operator's claims.
Concretely, this means that blockchain nodes would be assigned a new responsibility, that of producing a verifiable execution proof of their own code. This guarantees that participants cannot falsely declare the use of minority clients to gain undue rewards.

We realize our concept of verifiable client diversity in the concrete context of the Ethereum blockchain.
We consider the Lighthouse blockchain implementation for Ethereum, written in Rust. In Lighthouse, we single out functions that are amenable to be executed within two  state-of-the-art verifiable computation frameworks (zero-knowledge virtual machines and trusted execution environments).
We modify Lighthouse such that node operation always involve producing execution proofs of those function.
Next, we implement a smart contract that drives the reward mechanism. Blockchain nodes submit proofs of their identity to the contract, which keeps a tamperproof estimation of the distribution of each implementation accordingly. The smart contract is also responsible for distributing the rewards, in a transparent and verifiable way. To our knowledge,  this system is the the first ever of this kind being built end-to-end for a real blockchain, demonstrating the viability of the novel concept of verifiable client diversity.

The experimental evaluation focuses in determining the feasibility of our design by focusing on its effectiveness and performance.
For measuring its effectiveness, we deploy our protocol in a local Ethereum testnet, with perfectly rational, reward-maximizing node operators.
We observe that the distribution of client implementations evolves towards a uniform distribution.
Regarding performance, we measure and compare the time and resources used by the verifiable computation mechanisms.
We observe that there is a large tradeoff between the zkVM and TEE approaches, which is reflected in the time and resources it takes to produce proofs of execution with zkVMs, versus the cost and trust assumptions of TEE capable hardware.

To sum up, our contributions are:
\begin{itemize}
    \item \textbf{The novel concept of verifiable software diversity}: We introduce the innovative idea of verifiable software diversity, specifically tailored for blockchain networks. Verifiable Software Diversity provides guarantees on the resilience of a blockchain network by reducing the likelihood of systemic outages due to monoculture of blockchain nodes and supermajority clients.
    \item \textbf{Design of a verifiable client diversity protocol for blockchains}: We propose a detailed architectural blueprint for implementing verifiable client diversity in blockchain. This framework integrates economic incentives to encourage the adoption of minority clients and utilizes verifiable execution to corroborate the actual deployment and usage of client implementations.

    \item \textbf{Experimental evaluation}: We perform an end-to-end experimental evaluation results of our protocol, demonstrating feasibility, performance, and effectiveness, as well the major challenges given the verifiable computation state-of-the-art to date.

    \item \textbf{Prototype implementation of the proposed design}: We implement our prototype in the context of the Ethereum blockchain, exploring Trusted Execution Environments and zero-knowledge virtual machines as verifiable computation mechanisms. 
    Our prototype implementation is publicly available at \url{https://github.com/chains-project/verifiable-client-diversity/} for sake of future research on this topic.

\end{itemize}

\section{Background}

\subsection{Verifiable Computation}
Verifiable computation encompasses a family of cryptographic techniques that enable one party to prove to another that a computation was performed correctly, without requiring the verifier to re-execute the computation~\cite{survey-verifiable-computation}.
This field has evolved from simple proof systems to sophisticated protocols that can verify complex computations while maintaining security guarantees.
The key challenge in verifiable computation is to balance the computational overhead of proof generation and verification against the security guarantees provided.

In verifiable computation, we distinguish between two fundamental aspects:
(1) Code Integrity: the guarantee that the executed code matches a known, trusted implementation, and; 
(2) Execution Correctness: the assurance that the computation was performed faithfully according to the program's logic.

These aspects are tightly coupled through cryptographic primitives that bind proofs of execution to specific program implementations. This binding ensures that proofs can only be generated by running the exact intended code, preventing malicious actors from claiming to run some code while actually executing another.

The concept of code identity is central to verifiable computation.
We define \textit{code identity} as a cryptographic fingerprint of a program implementation that serves as a verifiable identifier.
This fingerprint is computed by applying a cryptographic hash function to the program that is to be verified, including its binary code, configuration, and critical runtime parameters.
The code identity enables us to distinguish between different implementations of the same functionality, even if they produce identical outputs.

The field has developed along two primary paradigms, each with distinct security assumptions and trade-offs:

\noindent\textbf{Trusted Execution Environment (TEE).} TEEs are secure areas of a processor that provide hardware-enforced isolation for code execution~\cite{TEE}.
Popular implementations include Intel SGX~\cite{sgx}, ARM TrustZone~\cite{trustzone}, and AMD SEV~\cite{SEV}.
TEEs achieve code integrity through hardware-level isolation mechanisms that prevent interference from the host OS or other processes.
The security of TEEs relies on the trustworthiness of the hardware manufacturer and the absence of critical vulnerabilities in the implementation.
To verify code execution, TEEs employ \textit{remote attestations}: cryptographically signed reports that identify the exact code, environment, and inputs for a program execution.
These attestations are verified using manufacturer-issued certificates and public keys, providing strong assurance of both code integrity and execution correctness.
However, TEEs face challenges such as side-channel attacks, limited memory enclaves, and the need to trust hardware manufacturers.

\noindent\textbf{Zero Knowledge Virtual Machine (zkVM).} The second paradigm uses cryptographic proofs, particularly zero-knowledge proofs, to attest computation correctness~\cite{zk-proof}.
In this approach, the prover generates a mathematical proof that the execution of a program on given inputs yielded a specific output.
Zero-knowledge VMs (zkVMs) are general-purpose virtual machines that produce proofs of execution~\cite{zkvm} alongside their outputs.
Modern zkVMs can prove the correct execution of general-purpose code (e.g., Rust or C code) using various proof systems, each with different trade-offs in terms of proof generation time, verification time, and proof size.
The key advantage of zkVMs over TEEs is that they don't require trust in hardware manufacturers.
However, they introduce significant computational overhead for proof generation and may have limitations in terms of the complexity of computations that can be proven efficiently~\cite{zkvm-perf}.

\subsection{Software Diversity for Reliability}
In the field of software engineering, \textit{software diversity} refers to the strategic creation and deployment of multiple functionally equivalent but internally different software versions~\cite{larsen2014sok}.
This approach enhances system security and reliability by mitigating risks associated with single points of failure through software faults or vulnerabilities in a monoculture.
By ensuring that not all components share the same vulnerabilities or fail in the same way, software diversity helps reduce the impact of failures or exploits~\cite{building}.

One of the classical approaches to building reliable systems using software diversity is N-version programming~\cite{avizheh2024refereed}, where multiple independent implementations of the same specification are developed and run in parallel.
These versions are designed to produce the same output given the same input, but due to being implemented by different teams, or using different design approaches~\cite{design-diversity}, programming languages~\cite{galapagos} or compilers~\cite{avatars}, they are less likely to fail in the same way.
This redundancy allows the system to detect and recover from faults by cross-verifying the outputs, typically using a majority voting mechanism~\cite{voting}.


In the context of distributed systems, software diversity is particularly valuable for improving fault tolerance and resilience~\cite{understanding, zhang2021chaoseth}.
Distributed protocols are often replicated across multiple nodes or services, and having several independent implementations of the same protocol can protect the system from uniform failure modes.
For example, if all nodes were to share the same bug due to a common codebase, a failure in one could propagate and affect the entire system, bringing down the entire distributed system.
However, with diverse implementations, such a bug is less likely to exist across all nodes, limiting the blast radius of any individual fault to the subset of nodes sharing the same implementation.

This is particularly relevant in security-critical infrastructures such as blockchain systems~\cite{proof-of-diversity} and cloud-native microservices~\cite{zheng2013selecting}, where resilience against both accidental faults and deliberate attacks is paramount.
In these contexts, software diversity must be carefully managed to ensure that different implementations remain functionally equivalent while maintaining their independence.

\section{Problem statement}
\label{sec:problem}

In this section, we examine two critical problems that arise from the lack of client diversity in blockchain networks: mass-slashing and network partitioning. These problems demonstrate how a skewed distribution of client implementations can lead to catastrophic failures, affecting both the network's security and its economic stability. We analyze each problem through both intuitive explanations and formal models to provide a comprehensive understanding of their impact.

\subsection{Client Diversity in Blockchain} 

Diversity in blockchains refers to the variety and distribution of the individual nodes that participate in maintaining the ledger. This diversity spans geographic locations, hardware configurations, client software implementations and node ownership. High diversity is crucial because it enhances the network's resilience to outages, censorship and coordinated attacks: if some nodes go offline or are maliciously targeted, others in different jurisdictions or running different software can keep the system in operation.

Client diversity, specifically, refers to the existence of multiple independent software implementations of a blockchain protocol. Its goal is to guard against bugs or backdoors that might be present in any single codebase. This is particularly important because different implementations, while functionally equivalent, are less likely to share the same vulnerabilities or fail in the same way. In the next section, we elaborate on two catastrophic scenarios caused by the lack of client diversity in blockchains: mass-slashing of nodes and chain forks.

\subsection{Problem 1: Mass-slashing of Blockchain Nodes} 

\begin{figure}[!t]
    \small
        \centering
        \includegraphics[width=0.95\textwidth]{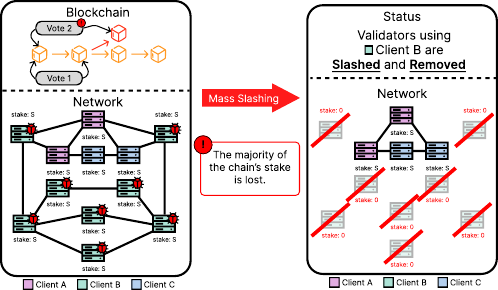}
        \vspace{3pt}
      \caption{A mass-slashing scenario due to lack of client diversity. A bug in \textit{ClientB} causes a majority of nodes to cast inconsistent votes in violation of consensus rules. Consequently, all users running \textit{ClientB} will be punished with a slashing penalty, and removed from the set of active validators. In the worst-case scenario \textit{ClientB} receives a penalty amounting to their full staked balance.}
      \label{fig_massiveslash}
\end{figure}

Consider a blockchain with a stake-based consensus, i.e. nodes are required to deposit an amount of native cryptocurrency in order to participate. One of the consensus' rules is that nodes must cast a vote for the one-and-only-one block to be appended to the chain. Penalties are applied to nodes who act negligently or maliciously, violating the consensus rules. In this case, a penalty can be applied to a node that votes simultaneously for more than one block to be appended to the chain.

In well-established consensus mechanisms such as Ethereum's Proof-of-Stake, such penalty is called \textit{Slashing}, and is essential for ensuring network integrity and security. Mass-slashing in blockchain refers to a situation where a large portion of the participants experience financial penalties at the same time~\cite{klin_post1,klin_post2,klin_post3}. 

\noindent\textbf{Intuition.} 
Fig.~\ref{fig_massiveslash} visualizes an instance of mass-slashing. In the figure, there are three types of clients in the blockchain network, \textit{ClientA} (pink), \textit{ClientB} (teal), and \textit{ClientC} (blue). Assume a bug in \textit{ClientB} that leads to inconsistent votes, which is a violation of a consensus rule. Therefore, all the users running \textit{ClientB} will get a penalty, being a mass slashing event in this example; 67\% of the network would disappear because of the bug in \textit{ClientB}.

\noindent\textbf{Formal model.} Assume a blockchain network with \(T\) nodes, where \(m\) clients implementations are deployed, denoted as subsets \(T_1, T_2, \ldots, T_m\). The cardinality \(|T|\) represents the total number of nodes within the blockchain network. Suppose \(T_1\) is the majority client, such that it satisfies \( \frac{1}{3} |T| \leq |T_1| \leq \frac{2}{3} |T| \).

Let the blockchain head be at height \(b_s\). Two subsequent blocks are proposed, \(b_{s+1}\) and \(b_{s+1}'\). Assume a majority client encounters a bug causing all nodes in \(T_1\) code to erroneously generate two votes: vote 1 from \(b_s\) to \(b_{s+1}\), and vote 2 from \(b_{s}\) to \(b_{s+1}'\). This violates the rules by making two differing attestations for the same block slot~\cite{ethdoc_penality}. In this context, mass-slashing means penalizing all nodes in \(T_1\), which is a large share of network participants.

\subsection{Problem 2: Corrupted Super-majority State.} 

In blockchain systems, maintaining a correct and consistent state across all nodes is fundamental to the network's operation. When a super-majority of nodes (more than 2/3 of the network) runs the same client implementation, a bug in that implementation can lead to catastrophic outcomes. Specifically, if the bug causes the super-majority to maintain an incorrect state, the network will effectively operate with this corrupted state since the super-majority's votes will be accepted by the consensus mechanism. This is particularly dangerous because the network will continue operating with incorrect state, potentially leading to invalid transactions being processed, incorrect account balances, or other critical ledger inconsistencies.

\begin{figure}[ht]
    \small
        \centering
        \includegraphics[width=0.95\textwidth]{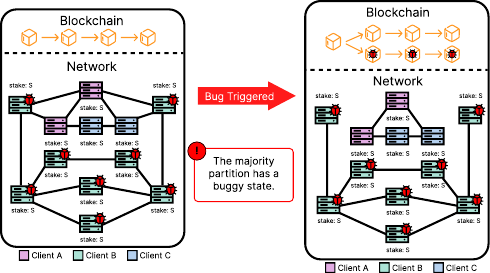}
        \vspace{3pt}
      \caption{A skewed distribution of clients compromises the blockchain consensus. In this example, the network consists of three client implementations: \textit{ClientA}, \textit{ClientB}, and \textit{ClientC}. Notably, \textit{ClientB} is the majority client, operating on eight out of twelve nodes. In the event of a bug affecting \textit{ClientB}, the blockchain can operate with an incorrect state since the majority of nodes will vote for invalid state transitions. This threatens the integrity and consistency of the blockchain.} 
      \label{fig_forkissue}
\end{figure}

\noindent\textbf{Intuition.}
Fig.~\ref{fig_forkissue} visualizes the risk of corrupted state due to a lack of client diversity. There are three types of clients in Fig.~\ref{fig_forkissue}, containing \textit{ClientA}, \textit{ClientB}, and \textit{ClientC}. \textit{ClientB} is the majority client, as eight of the twelve nodes operate it. When \textit{ClientB} experiences a bug that corrupts its state, the network will follow this incorrect state since the majority of nodes will vote for invalid state transitions. 

\noindent\textbf{Formal model.}
Assume a blockchain network with \(T\) nodes, where \(m\) types of blockchain clients are deployed, denoted as subsets \(T_1, T_2, \ldots, T_m\). The cardinality \(|T|\) represents the total number of nodes within the blockchain network. Each blockchain node maintains a local state \( S_T \).

In a normal scenario where consensus is maintained, we have \( \forall t_i, t_j \in T, S_{T_i} = S_{T_j} \), indicating that all nodes keep the same blockchain state. Assume that implementation $T_2$ is a super-majority client, with \( |T_2| > 2/3\times|T| \). If a bug is triggered in \(T_2\) that corrupts its state, the nodes in \(T_2\) will maintain an incorrect state \(S_{T_2}^{incorrect}\), while nodes in other implementations maintain the correct state \(S_{T_2}^{correct}\). Due to the super-majority status of \(T_2\), the network will effectively operate with the incorrect state, as \( \forall t_i \in T_2, S_{T_i} = S_{T_2}^{incorrect} \) and this state is accepted by the consensus mechanism.

In this paper, we propose a systemic approach to mitigate the fundamental problems resulting from a monoculture in a blockchain system.

\section{System Design for Verifiable Client Diversity}

Our system design aims to create a robust framework that actively promotes and maintains client diversity in blockchain networks.
The key insight is that tamperproof economic mechanisms can ensure the actual adoption and usage of diverse implementations.

This chapter describes the design of our proposed system to achieve client diversity through two key mechanisms: Client Implementation Proof-of-Execution and Financial Incentives for Minority Clients.
Together these mechanisms form an end-to-end client diversity protocol: a verifiable framework for encouraging heterogeneous client usage by integrating proof-of-execution verification with on-chain financial incentives.
This design naturally creates a self-stabilized system where diversity is achieved and maintained, with provable means.

The high-level workflow of our system is as follows:
We combine proofs-of-execution and financial incentives in two stages:
(1) a commitment stage, where each client implementation's code is processed to create cryptographic commitments that serve as unique code identities, and;
(2) a runtime stage, where nodes generate proofs of execution that can be verified against the stored commitments to confirm which client implementation they are running.
The commitments are computed once and stored on-chain, the proofs are generated and submitted during normal blockchain operation, and verified on chain.

        



\subsection{Commitment to Code Identity}
\label{client-implementation-verification}

The objective of this stage is to produce verifiable, tamper-resistant facts about the code identity of participants in the blockchain network.

In this first stage, each client implementation's code is processed to create \textit{cryptographic commitments} that serve as unique code identities.
For this, the protocol leverages existing technology, such as zero-knowledge execution or trusted execution environment.
The commitments are computed once and stored on-chain.

Creating a proof of execution for an entire, continuously-running blockchain client is impractical with off-the-shelf technology, because it would require proving the execution of millions of lines of code across multiple processes, with complex state transitions and external interactions.
To work around this fundamental limitation, we propose to approximate full code identity by considering a key specific functionality to be proven.

Consider a blockchain protocol $P = \{~p_0,~p_1,~...,~ p_n~\}$, where every $p_x \in P$ is a specific, well-defined step of $P$ performed for each block.
Practical examples of $p_x$ are: the aggregation of transactions into a block; the signature of an assembled block; or the verification of the block well-formedness per the protocol rule.

Consider the set of client implementations $I = \{~I_0,~I_1,~...~,~I_m~\}$, where each distinct client implementation $I_y \in I$ is a program which complies with protocol $P$.
The subset of instructions $ A =\{~p^{I_0}_x,~p^{I_1}_x,~...,~ p^{I_m}_x~\}$ is the implementation of $p_x$ for each client implementation, where every $p^{I_y}_x$ is distinct enough to serve as a reliable fingerprint of $I_y$.

Using a verifiable computation mechanism, we can compute a set of program commitments $C = \{~C_0,~C_1,~...,~C_m~\}$, where every $C_y \in C$ corresponds to each element in $A$.
These commitments are used to verify the usage of any implementation $I_y$ for the execution of protocol step $p_x$ at any block.
\autoref{alg:commitment} summarizes the commitment to code identity stage.

\begin{algorithm}[H]
\begin{algorithmic}[1]
\Require Set of client implementations $I = \{I_1, I_2, ..., I_n\}$
\Require Set of protocol steps $P = \{p_1, p_2, ..., p_m\}$
\For{each client implementation $I_y \in I$}
    \For{each selected protocol step $p_x \in P$}
        \State Extract code segment $p_x^{I_y}$ implementing $p_x$ in $I_y$
        \State Compute cryptographic commitment $C_{y,x} = \text{Hash}(p_x^{I_y})$
    \EndFor
\EndFor
\State Store all commitments $C_{y,x}$ on-chain as the set of valid code identities
\Ensure On-chain registry of commitments $\{C_{y,x}\}$ uniquely identifying each client implementation and protocol step
\end{algorithmic}
\caption{Commitment to Code Identity}
\label{alg:commitment}
\end{algorithm}

\subsection{Proof Production Stage}

The proof production stage happens during normal blockchain operation, for each block.
The concept is that nodes regularly generate and submit proofs of execution to provably demonstrate their use of specific client implementations.
Let $N = \{n_1, n_2, ..., n_k\}$ be the set of nodes in the network, where each node $n_i \in N$ runs a client implementation $I_y \in I$ from the set of approved implementations.

For a critical operation $p_x \in P$ performed by node $n_i$, a proof $\pi_{n_i}^{p_x}$ is generated. The proof $\pi_{n_i}^{p_x}$ cryptographically attests to the execution of code matching commitment $C_y \in C$, ensuring that the node is running the claimed implementation. 

Each node $n_i$ submits a tuple $(b, \pi_{n_i}^{p_x})$ to the smart contract $T$, where $b$ is the block number, and $\pi_{n_i}^{p_x}$ is the proof of execution. 

To sum up, the core novelty of our design is that blockchain nodes execute some of their functions in a provable way.
This opens the door for many key innovations, including verifiable client diversity.

\begin{figure}[h]
    \small
        \centering
        \includegraphics[width=0.65\textwidth]{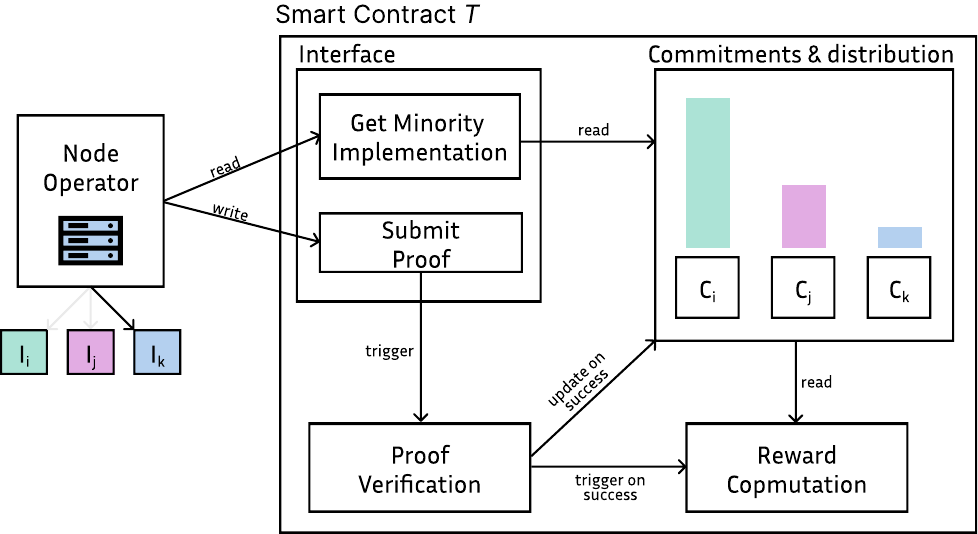}
        \vspace{3pt}
      \caption{A smart contract stores a tamper-proof estimation of the current node implementation distribution. For every block, node operators can query the contract to select the implementation to run in order to maximize their rewards.} 
      \label{fig:model-2}
\end{figure}

\subsection{Client Diversity Reward through Smart Contracts}
\label{sec:protocol}
The last part of the design is the verification of the submitted proofs and the computation of the corresponding reward.
This happens through a protocol implemented as a smart contract $T$.

Each time an execution proof is submitted, the smart contract verifies it using an on-chain implementation of the verifier.
Upon successful verification, the smart contract $T$ updates the current distribution of client implementations, calculates the reward $r$ based on the implementation's current share, and transfers $r$ to the proof submitter.
\autoref{fig:model-2} visualizes the model described. The node operator is shown interacting with the smart contract to submit a proof of execution for a specific step $p_x$ of the protocol.
The smart contract verifies the proof and updates the distribution of client implementations.

Contract $T$ is initialized with:
(1) a set of commitments $C$, those accepted as valid code identities in the protocol (the selection of $C$ is a governance decision discussed in \autoref{sec:governance});
(2) a constant reward $\epsilon$;
(3) a maximum additional reward $r_{max}$; and 
(4) a minimum additional reward $r_{min}$.
As for its interface, $T$ exposes two functions: one function for nodes to submit and apply for reward  $submit\_proof$; and one function to query the current minoity implementation $get\_minority$.

Function $submit\_proof$ takes a proof $\pi$ as a parameter, which must match a commitment $C_y \in C$.
If the proof is successfully verified, it means that the user has demonstrated genuine use of an approved client implementation for step $p_x$.
The function then assigns rewards to the proof submitter with $r$, as described below.
Finally, $submit\_proof$ updates the client distribution by incrementing the corresponding counter.

Function $get\_minority$ takes no parameters, and returns the current minority implementation, allowing the user to know with confidence which client implementation is the least used and gives the highest reward.

Reward $r$ is computed as a function of the submitted commitment $C_y$ and the current distribution, in order to aim at a uniform distribution over client implementations.
Let $|~C~|$ be the cardinality of $C$.
If commitment $C_y$ represents more than $1~/~|~C~|$ of the distribution, the reward is $\epsilon$.
If the proof $C_y$ is submitted by an operator who is not expected for the given block, the reward is zero. 
If commitment $C_y$ represents less or equal to $1~/~|~C~|$ of the distribution, the reward is computed as a linear interpolation between $r_{max}$ and $r_{min}$, where the interpolation parameter is $C_y*|~C~|$.

\autoref{fig:reward-function} visualizes this reward function. 
The protocol: 
(1) provides the maximum reward to proofs that match the commitment that has been used the least (left-most point);
(2) provides rewards to proofs to users that contribute to improving client diversity (green part);
(3) provides minimal rewards to users executing a majority client, making it attractive to report their information, even if not willing to switch to a minority client (red part).
More sophisticated functions can be designed and plugged into the core protocol.

\begin{figure}[h]
    \centering
    \begin{tikzpicture}[scale=0.8]
        \fill[color1bg!50] (0,0) rectangle (5,7);
        \fill[color2bg!50] (5,0) rectangle (10,7);
        
        \draw[gray, thin] (0,0) grid (10,7);
        
        \draw[line width=0.5mm, color1line] (0,6) -- (5,2) -- (5,1) -- (10,1);
        
        \node[below] at (5,0) {$\frac{1}{|C|}$};
        \node[below] at (10,0) {1};
        \node[left] at (0,6) {$r_{max}$};
        \node[left] at (0,2) {$r_{min}$};
        \node[left] at (0,1) {$\epsilon$};
        \node[below] at (0,0) {0};
        
        \draw[dashed] (5,1) -- (5,6);
        \draw[dashed] (0,6) -- (5,6);
        \draw[dashed] (0,2) -- (5,2);
        \draw[dashed] (0,1) -- (5,1);
        
        \node[align=center] at (2.5,8.5) {Linear Interpolation\\$r_{max}$ to $r_{min}$};
        \node[align=center] at (7.5,8.5) {Constant $\epsilon$ reward};
        \node[align=center] at (2.5,-1) {Minority Clients};
        \node[align=center] at (7.5,-1) {Majority Clients};

        \draw[->] (0,0) -- (10,0) node[right] {Client Share};
        \draw[->] (0,0) -- (0,7) node[above] {Reward ($r$)};

    \end{tikzpicture}
    \caption{Reward function visualization for the client diversity protocol. The reward starts at $r_{max}$ for share 0 and linearly decreases to $r_{min}$ as the client share approaches $\frac{1}{|C|}$. An $\epsilon$ reward is given for majority client in order to encourage participation.}
    \label{fig:reward-function}
\end{figure}

\autoref{alg:proof-reward} summarizes the proof production and diversity reward stages.

\begin{algorithm}[H]
\caption{Proof Production and Diversity Reward Protocol}
\label{alg:proof-reward}
\begin{algorithmic}[1]
\Require Set of nodes $N = \{n_1, n_2, ..., n_k\}$
\Require Set of approved client implementations $I$ and commitments $C$
\Require Protocol steps $P$
\Require Smart contract $T$ with reward parameters
\For{each block}
    \For{each node $n_i \in N$}
        \State Select protocol step $p_x \in P$ to execute
        \State Execute $p_x$ using client implementation $I_y$
        \State Generate proof of execution $\pi_{n_i}^{p_x}$ attesting to code matching commitment $C_y$
        \State Submit tuple $(b, \pi_{n_i}^{p_x}, addr_{n_i})$ to smart contract $T$
    \EndFor
    \For{each submission $(b, \pi_{n_i}^{p_x}, addr_{n_i})$ received by $T$}
        \If{$\pi_{n_i}^{p_x}$ does not verify against $C$}
            \State Reject submission
        \Else
            \State Update distribution of client implementations
            \If{share of $C_y \leq 1/|C|$}
                \State Compute reward $r$ as linear interpolation between $r_{max}$ and $r_{min}$
            \Else
                \State Set reward $r = \epsilon$
            \EndIf
            \State Transfer reward $r$ to $addr_{n_i}$
        \EndIf
    \EndFor
\EndFor
\Ensure Updated on-chain distribution and rewards distributed according to diversity protocol
\end{algorithmic}
\end{algorithm}

From an economic perspective, this mechanism aligns incentives with the goal of client diversity:
by biasing rewards toward minority clients, the distribution is driven toward uniformity in the long run.
In this way, the smart contract enforces transparent, automated, and trustless incentives that promote a balanced ecosystem of client implementations.
Under no friction for switching, rational users will tend to submit a proof if it is profitable to do so.
We give more details about implementation for Ethereum in section~(\autoref{sec:implementation}).

\subsection{System Properties}

The proposed system directly addresses the critical problems of monoculture and super-majority clients identified in \autoref{sec:problem} by creating an economic incentive around client implementation choices.
By providing higher rewards for minority clients and zero rewards for majority clients, the system creates a natural pressure against the formation of super-majority clients that could result in mass-slashing events or irreconcilable network partitions.

\noindent\textbf{Economic Properties.}
The system's economic design ensures several key properties. \revised{The reward mechanism naturally drives the network toward a balanced distribution of client implementations, as nodes are incentivized to switch to minority clients when they become profitable to operate. Once a balanced distribution is achieved, the system maintains stability as the economic incentives prevent any single client from gaining a super-majority position. The reward function automatically adjusts based on the current distribution, ensuring that incentives remain effective in any network state.}

\noindent\textbf{Resilience Properties.}
The system enhances blockchain resilience through several key mechanisms. First, it maintains a balanced distribution of client implementations through economic incentives, reducing the impact of bugs or vulnerabilities in any single client. Second, it provides strong guarantees against manipulation through the combination of economic incentives and verifiable execution, making it economically infeasible to manipulate the client distribution through false claims. Third, in the event of a critical bug affecting a share of the clients, the system's economic incentives can be quickly updated to encourage nodes to switch to unaffected clients, facilitating network recovery. These mechanisms work together to create a self-stabilizing system that maintains resilience against both accidental faults and deliberate attacks.

\noindent\textbf{Operational Properties.}
The system maintains several important operational characteristics. 
The system is fully verifiable: all aspects of the system, including client distribution, rewards, and verification processes, are publicly verifiable on the blockchain. 
The protocol is fully automated: the entire process of proof generation, verification, and reward distribution operates automatically without requiring manual intervention. 
The protocol is extensible: the system can accommodate new client implementations without requiring changes to the core protocol, allowing for organic growth of the client ecosystem.

These properties collectively ensure that the system achieves the primary goal of promoting and maintaining client diversity while remaining secure, resilient, and economically sustainable.

\section{Implementation}
\label{sec:implementation}


In this section we describe a prototype implementation of the proposed system, in the context of the Ethereum blockchain.

The verifiable computation component is implemented as a module supporting both zkVM and TEE approaches to demonstrate the system's flexibility and evaluate different tradeoffs in practice.
Specifically, we use RISC Zero and Intel SGX, respectively, as representative implementations of these approaches.
For the client diversity protocol, we implement our design in a Solidity smart contract.

\subsection{Proof of Execution for Ethereum Clients}

We selected Lighthouse as our target client node for implementation.
Running the entire Lighthouse codebase in a zkVM or TEE would be computationally infeasible due to its size and complexity.
Instead, we carefully analyzed the codebase to identify critical functions that are both essential for client operation and feasible to verify.
After analyzing the full code base of Lighthouse, we have selected the functions shown in \autoref{tab:functions}, focusing on key operations like attestation creation and block signing that are fundamental to client operation while being computationally tractable for verification.

We selected Lighthouse as our target client node for implementation.
Running the entire Lighthouse codebase in a zkVM or TEE would be computationally infeasible due to its size and complexity.
Instead, we carefully analyzed the codebase to identify critical functions that are both essential for client operation and feasible to verify.
After analyzing the full code base of Lighthouse, we have selected the functions shown in \autoref{tab:functions}.
They focus on key operations like attestation creation and block signing that are fundamental to client operation.
Both of them computationally tractable for verifiable computation with the considered technologies.
For example, the \texttt{create\_unaggregated\_attestation} function is used to create attestations for new blocks, which is a critical operation for maintaining consensus.
This function is provably executed using RISC Zero to ensure the attestation was created correctly.

\begin{table}
    \centering
    \begin{tabular}{llll}
        \toprule
        \textbf{ID} & \textbf{Function} & \textbf{Protocol step} & \textbf{Proving method} \\
        \midrule
        F1 & \small\texttt{beacon\_chain::create\_unaggregated\_attestation} & Block attestation & zkVM \\
        F2 & \small\texttt{crypto/bls::SecretKey.sign()} & Block signature & TEE \\
        \bottomrule
    \end{tabular}
    \caption{Real blockchain node functions target for verifiable computation. We prototype with two proving methods, zero-knowledge VMs and trusted execution environments. Our proof-of-concept prototype is built on top of the Lighthouse client implementation of Ethereum.}
    \label{tab:functions}
\end{table}

\noindent\textbf{Proof of Execution Using zkVM.}
The zkVM approach involves three key steps. First, we isolate the code of a function from \autoref{tab:functions} in a separate Rust module.
We use \href{https://github.com/risc0/risc0/tree/v1.2.1}{RISC Zero 1.2.1} to compile the module to RISC-V.
The program commitment $C^{zk}_{fn}$ uniquely identifies the compiled code.
At runtime in the Ethereum node, the execution of $p_x$ is done within RISC Zero's virtual machine, generating a succinct, cryptographic proof.
Finally, this proof is submitted to the smart contract, which uses RISC Zero's publicly available verification contracts to verify the proof's validity.

\noindent\textbf{Proof of Execution Using TEEs.}
The TEE approach also follows three key steps. First, we use Intel SGX, isolate the code of a function from \autoref{tab:functions} in a TEE \textit{enclave} using the \href{https://github.com/automata-network/automata-sgx-sdk}{Automata SGX SDK}.
This allows us to compute the program commitment $C^{tee}_{fn}$.
Next, we run the function in the TEE enclave.
Upon successful execution, the TEE enclave issues a cryptographically signed data structure as proof indicating that the specified function was correctly executed.
Finally, this proof is submitted to the smart contract, which verifies it using manufacturer-issued certificates and public keys, thus providing strong assurance of both code integrity and execution correctness.
\begin{figure}[h]
    \centering
    \includegraphics[width=0.95\linewidth]{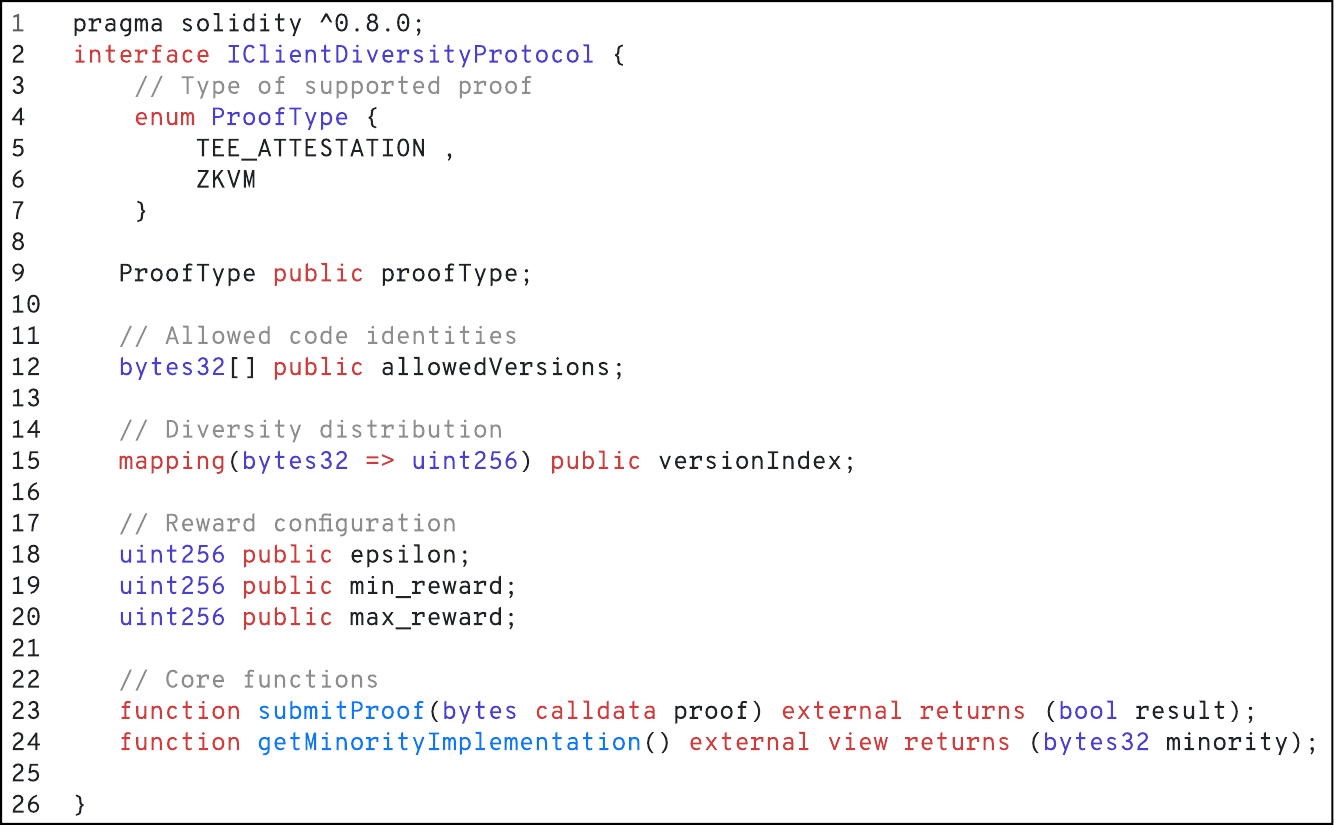}
    \caption{Client Diversity Protocol Smart Contract Interface.}
    \label{lst:spec}
\end{figure}

\subsection{Client Diversity Protocol}

We implement the rewarding mechanism described in \autoref{sec:protocol} as a smart contract in Solidity, to be deployed on Ethereum or any EVM compatible chain.
This smart contract supports verification of SGX and RISC0 proofs against the registry of owner-approved program commitment hashes.
The public API exposes methods to retrieve supported versions, submit proofs, and query the current minority version, while the owner—guarded modifiers may adjust reward amounts, and manage approved versions.
\autoref{lst:spec} Shows the specification of the smart contract. The full implementation is available in \href{https://github.com/chains-project/verifiable-client-diversity/blob/main/spec.md}{our GitHub repository.}

\section{Experimental Evaluation}
\subsection{Research Questions}
To evaluate our concept of verifiable client diversity, we study the following research questions.
\begin{enumerate}
    
    \item[\textbf{RQ1}] \textit{What is the execution overhead introduced by proving client node execution?}

    \vspace{7pt}
    Verifiable computation does not come for free.
    We measure and report the resources used to produce proofs in the context of blockchain client diversity. \\
    
    \item[\textbf{RQ2}] \textit{What is the cost of verifying provable execution of client implementations on chain?}

    \vspace{7pt}
    On-Chain verification of provable computation does not come for free.
    We measure and report the gas cost to verify execution proofs in the context of our client diversity protocol.
    This is crucial, since the reward must be higher than the cost of verifying it.\\
    
    \item[\textbf{RQ3}] \textit{To what extent does the client diversity protocol incentivize blockchain client diversity?}

    \vspace{7pt}
    We aim to capture the system-wide behavior of blockchain operators when client diversity incentives are introduced. Specifically we show how implementation distribution changes over time, given potential rewards for using tamperproof minority implementations.\\

\end{enumerate}

\subsection{Methodology for RQ1 (Proof Production)}

The protocol has different components, with proof production critically impacting the practical feasibility of our system.
To evaluate the performance overhead of our verifiable execution mechanisms, we conduct a comprehensive performance analysis of both zkVM and TEE approaches.
Our methodology consists of three main components: (1) test environment setup, (2) performance metrics collection, and (3) comparative analysis.

\noindent\textbf{Test Environment Setup.}
We establish two distinct test environments to evaluate each proving mechanism:

\begin{itemize}
    \item \textbf{zkVM Environment:} We use a high-performance workstation with an Intel i9 processor, 128GB RAM, and 32 cores to produce RISC Zero execution proofs. This configuration ensures we can measure the upper bounds of zkVM performance.
    
    \item \textbf{TEE Environment:} We deploy our TEE-based implementation on Azure's SGX-enabled virtual machines, which provide the necessary hardware support for Intel SGX attestations.
\end{itemize}

\noindent\textbf{Performance Metrics Collection.}
For each function listed in \autoref{tab:functions}, we measure the total time required to generate a proof, CPU usage (percentage and core distribution), memory consumption (peak and average), and proof size. These metrics provide a comprehensive view of the performance characteristics of each proving mechanism.
We conduct multiple test runs to ensure statistical significance and account for network conditions.

\noindent\textbf{Comparative Analysis.}
We analyze the performance data to compare overhead between zkVM and TEE approaches, providing insights into the practical limitations and trade-offs of each method.

\subsection{Methodology for RQ2 (Proof verification)}

The protocol is based on on-chain verification, which means that all proofs must be verified by a smart contract.
This verification process incurs gas costs that must be carefully considered in the economic design of the system.
To evaluate the on-chain verification costs of our client diversity protocol, we conduct a detailed gas cost analysis.
Our methodology consists of two main components: verification cost measurement and economic analysis.

\noindent\textbf{Verification Cost Measurement.}
We deploy our smart contract on a testnet and measure gas costs for both proof verification and protocol operations. For proof verification, we track zkVM proof verification costs and TEE attestation verification costs. For protocol operations, we monitor distribution updates and reward calculations.

\noindent\textbf{Economic Analysis.}
We analyze the economic viability of the protocol by comparing verification costs against potential rewards and calculating break-even points for different client distributions.

\subsection{Methodology for RQ3 (Protocol Properties)}

To evaluate the effectiveness of our client diversity protocol, we conduct a controlled, end-to-end experiment that simulates the behavior of rational validators in a testnet environment. Our methodology consists of three main components: (1) implementation diversity simulation, (2) validator behavior modeling, and (3) distribution analysis.

\noindent\textbf{Implementation Diversity Simulation.}
We create variants of the Lighthouse client implementation by modifying the functions listed in \autoref{tab:functions}, such that they produce distinct program commitments. Each modification yields in a unique code identity while maintaining functional equivalence. This approach allows us to simulate multiple client implementations without running full clients. The modifications are designed to be indistinguishable from real client diversity in terms of their cryptographic properties and proof generation.
In total, we create 3 synthetic Lighthouse variants with distinct, provable code identity.

\noindent\textbf{Validator Behavior Modeling.}
We deploy a testnet with 12 validators that follow a rational economic strategy: they always choose to run one of the 3 client implementations, in order to maximize their rewards. Each validator continuously monitors the current distribution of client implementations through the smart contract, switches between different implementations if need be, and generates and submits proofs of execution for their chosen implementation to receive rewards based on the current distribution.

\noindent\textbf{Distribution Analysis.}
We conduct two distinct experimental scenarios to evaluate the protocol's behavior under the following conditions.
We initialize the network with three client implementations (A, B, and C) where implementation A starts with a supermajority (83\% of validators), and B and C 8.3\% each. This scenario tests the protocol's ability to naturally balance the distribution through economic incentives.

We collect and analyze the following metrics over time:
\begin{itemize}
    \item Distribution of client implementations among validators
    \item Time taken to reach equilibrium (if achieved)
    \item Stability of the distribution once equilibrium is reached
\end{itemize}

We use these metrics to visualize the evolution of client implementation distribution and the corresponding reward levels.
This allows us to assess both the effectiveness of the economic incentives and the protocol's ability to maintain a balanced distribution of client implementations.
\section{Experimental Results}

\subsection{RQ1 Proof Production Overhead}

\begin{table}[h]
    \centering
    \begin{tabular}{ll rrr}
        \toprule
        Function & Proof Type & Proof Generation Time & Regular Execution Time & Overhead  \\
        \midrule
        F1 & zkVM & \numprint{59} s & 15.14 $\mu$~s & \numprint{39333333}x  \\
        F2 & TEE & \numprint{80} ms & \numprint{1.42} ms & \numprint{56}x  \\
        \bottomrule
    \end{tabular}
    \caption{Execution time overhead for proof production by proof type mechanism. zkVM execution proofs are attractive in theory but infeasible in practice within blockchain nodes.}
    \label{tab:proving-overhead}
\end{table}

\begin{table}[h]
    \centering
    \begin{tabular}{ll rrrr}
        \toprule
        Function & Proof Type & CPU avg. & CPU max. & Mem. avg. & Mem. max. \\
        \midrule
        F1 & zkVM & \numprint{90.05}\% & \numprint{100.00}\% & \numprint{1331}MB & \numprint{2150}MB \\
        F2 & TEE  & \numprint{22.24}\% & \numprint{23.35}\% & \numprint{21}MB & \numprint{21}MB \\
        \bottomrule
    \end{tabular}
    \caption{Resource consumption of proof production by proof mechanism, confirming the intractability of zkVM proofs for blockchain nodes. }
    \label{tab:proving-resources}
\end{table}

Table~\ref{tab:proving-overhead} presents a detailed breakdown of the overhead associated with proof production for two critical functions in the Lighthouse Ethereum client, using both zkVM and TEE mechanisms. For the function F1, which utilizes zkVM, the proof generation time is 59 seconds, compared to a regular execution time of just 15.14 microseconds, resulting in an overhead of approximately 39 million times. In contrast, function F2, proven using a TEE, exhibits a much lower proof generation time of 80 milliseconds versus a regular execution time of 1.42 milliseconds, corresponding to a 56-fold overhead. This stark difference highlights the substantial computational cost of zkVM-based proofs relative to TEE-based execution proofs.

Resource consumption metrics, shown in Table~\ref{tab:proving-resources}, further underscore these differences. For zkVM (F1), the average CPU usage during proof generation is 90.05\%, peaking at 100\%, with memory usage averaging 1331MB and reaching a maximum of 2150MB. These results confirm that zkVM proof generation is highly resource-intensive, both in terms of computation and memory. The process is compute-bound, saturating available CPU resources, and may not be practical for resource-constrained blockchain nodes.
Also, it clearly shows the infeasibility of running the full node with a zk virtual machine.

In comparison, the TEE-based approach (F2) is significantly more efficient, with average and peak CPU usage at 22.24\% and 23.35\%, respectively, and memory usage remaining constant at 21MB.  TEEs, by leveraging hardware-based attestation provide faster, more efficient proof production but with different trust assumptions. 


When comparing the average proof generation times to standard Ethereum block times, typically 12 seconds per block, it becomes clear that zkVM-based proof generation for F1 (59 seconds on a powerful machine) far exceeds the block interval, making it impractical for real-time or per-block proofs in its current form. In contrast, TEE-based proofs for F2 are well within the block time, suggesting that TEE attestation is more suitable for real world deployment in general and client diversity proving in particular. 

\begin{tcolorbox}[title=Answer to RQ1]
Our experimental results show that zkVM-based proof generation introduces an extremely high computational overhead, up to 39 \textit{million} times slower than regular execution.
zkVM requires significant CPU and memory resources and taking nearly a minute per proof for a single protocol step, which is impractical for real-time blockchain operation.
In contrast, TEE-based proof generation is much more efficient, with proof times being within typical block intervals, making it feasible for real-world deployment in blockchain nodes.
These findings indicate that, while zkVMs offer strong security guarantees, their current performance limitations make TEEs the more practical choice for implementing verifiable client diversity in real-world blockchain systems.
\end{tcolorbox}

\subsection{RQ2 Proof Verification Overhead}

\begin{table}[h]
    \centering
    \begin{tabular}{ll rrr}
        \toprule
        Function & Verification Contracts & Gas usage min. & Gas usage avg. & Gas usage max. \\
        \midrule
        F1 & \href{https://github.com/risc0/risc0-ethereum/tree/main/contracts/src}{RISC Zero verifier} & \numprint{288458} & \numprint{289728} & \numprint{291013} \\
        F2 & \href{https://github.com/automata-network/automata-dcap-attestation/tree/main/evm/contracts}{Automata DCAP Attestation} & \numprint{5397746} & \numprint{5397746} & \numprint{5397746} \\
        \bottomrule
    \end{tabular}
    \caption{Gas usage for proof verification by function and mechanism. TEE proof verification costs have to be offset by the reward in order to be economically viable.}
    \label{tab:verification-overhead}
\end{table}

Table~\ref{tab:verification-overhead} presents the gas usage required for on-chain verification of execution proofs, comparing the two considered mechanisms: zkVM-based proofs (RISC Zero) and TEE-based proofs (Intel SGX). The results reveal a dramatic disparity in verification costs between these approaches. For the zkVM-based function (F1), the gas usage for proof verification is remarkably low, with an average of approximately \numprint{289728} gas units and only minor variation between minimum and maximum observed values. This efficiency is a testament to the maturity of zero-knowledge proof verification circuits, which are designed to minimize computational burden despite the complexity of the underlying computation that is being verified.

In stark contrast, the TEE-based function (F2) incurs a much higher and constant gas cost of \numprint{5397746} gas units for each verification. This substantial overhead reflects the additional steps required to validate hardware-based attestations, such as signature checks, certificate chain validation, and more complex data structure handling. The lack of variance in the TEE verification cost is likely due to a deterministic verification path with fixed-size cryptographic operations rather than variable proof content.

The implications of these experimental results are significant for protocol design. 
zkVM-based proofs are computationally expensive to generate off-chain (see RQ1), but their on-chain verification is highly efficient and scalable. This makes them attractive for cases where one produces proofs rarely but verifies them frequently. However, this is not the profile of our client diversity protocol.
TEE-based proofs are faster and cheaper to produce off-chain, but impose a heavy on-chain cost that could limit their practicality in high-throughput environments or when gas prices are volatile.
Overall, TEE-based proofs meet more constraints that are key to the proposed client diversity protocol, and are more suitable for the proposed protocol.

These results motivate further research into optimizing both proof generation and verification, as well as exploring alternative mechanisms that can deliver strong code diversity guarantees without prohibitive on-chain or off-chain costs.

\begin{tcolorbox}[title=Answer to RQ2]
The results demonstrate that verification of zkVM-based proofs in smart contracts is efficient, requiring less than \numprint{300000} gas per verification, which is suitable for frequent verification.
In contrast, TEE-based proof verification is significantly more expensive, consuming over \numprint{5000000} gas per verification, which could be prohibitive in high-throughput or high-gas-price environments.
While zkVMs are very costly to use off-chain, their on-chain verification is practical; TEEs, though efficient off-chain, impose a heavy on-chain cost. 
Overall, TEE-based proofs are more suitable for a verifiable client diversity protocol.
\end{tcolorbox}

\subsection{RQ3 Protocol Behavior}

To evaluate the real-world effect of the client diversity protocol, we conducted a controlled experiment running 12 rational validators and 3 client implementations (A, B, and C, all Lighthouse variants).
The initial distribution is highly skewed: implementation A held a supermajority: 70\% of validators, while B and C each started with only 20\% and 10\% respectively. Validators are programmed to be rational to always select the client implementation that maximizes their expected reward, dynamically switching in response to the current distribution as tracked by the smart contract.

\begin{figure}[b]
    \centering
    \includegraphics[width=0.95\linewidth]{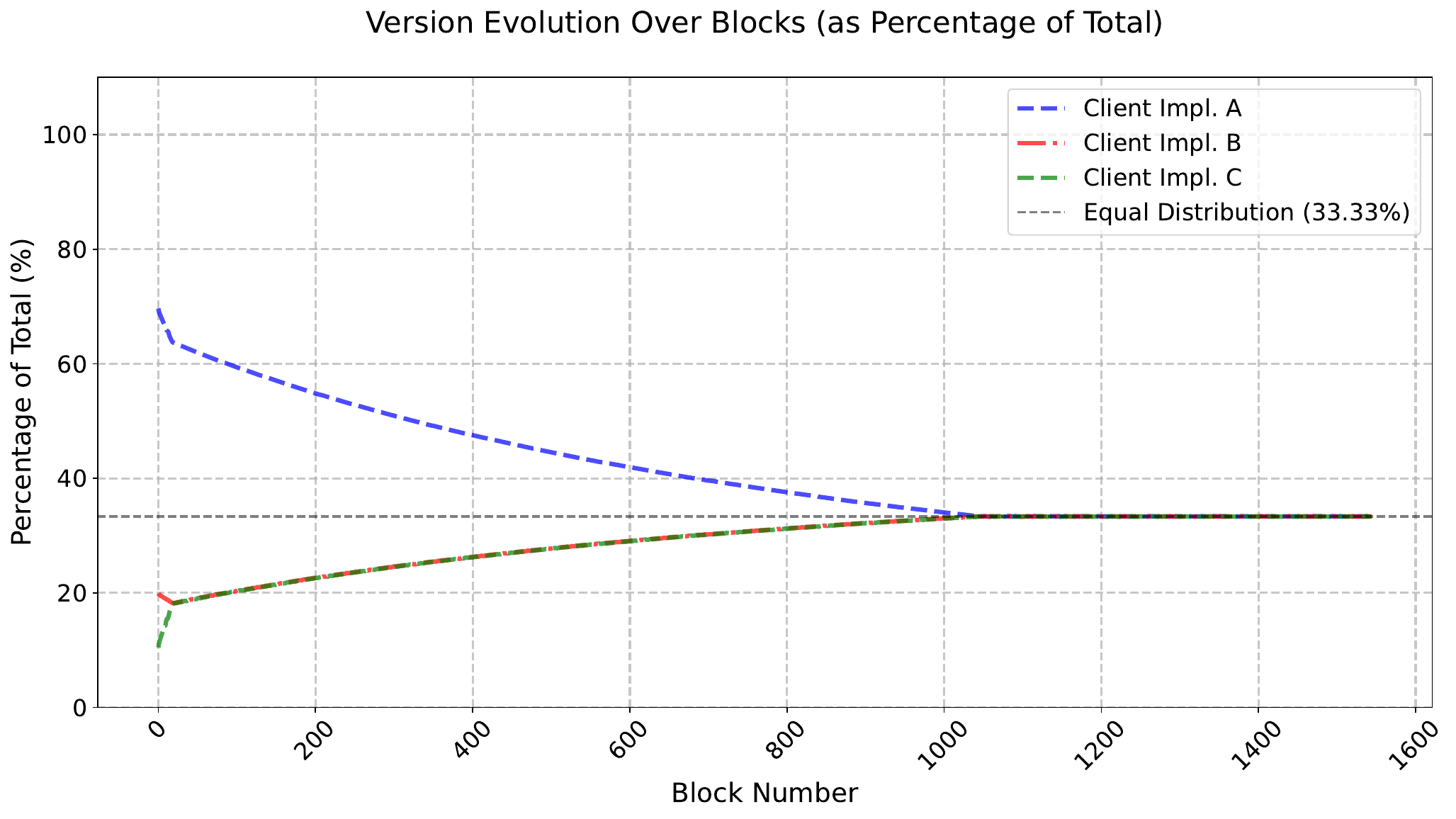}
    \caption{Evolution of client implementation distribution over time. The protocol drives the network from an initial supermajority (A: 70\%, B: 20\%, C: 10\%) to a uniform distribution, demonstrating convergence and stability of client diversity incentives. The final, stable, uniform distribution minimizes systemic weaknesses leading to catastrophic events, as the ones described in Section \ref{sec:problem}.}
    \label{fig:rq1-results}
\end{figure}

Figure~\ref{fig:rq1-results} shows the evolution of client implementation distribution over time, as validators respond to the protocol's economic incentives. The protocol rapidly drives the network away from a supermajority: the share of implementation A decreases as validators migrate to the minority clients B and C, whose shares increase. Over the course of several hundred blocks, the distribution converges to a near-uniform state, with all three implementations being equally represented. Once equilibrium is reached, the distribution remains stable, demonstrating the protocol's effectiveness in maintaining client diversity.
It is important to note that the number of blocks required for the validator distribution to reach equilibrium varies depending on the initial conditions. For example, a more extreme initial supermajority or a greater number of client implementations can increase the time required for the protocol to achieve a uniform distribution. The convergence speed is influenced by both the initial skew and the responsiveness of validators to incentive changes. 

This experiment confirms that the proposed reward mechanism successfully incentivizes validators to adopt minority clients, eliminating supermajority risk and promoting a balanced, resilient network. The total rewards distributed follow the protocol's rules, with higher rewards going to validators running minority clients during the convergence phase.

\begin{tcolorbox}[title=Answer to RQ3]
The results show that our proposed client diversity protocol effectively incentivizes node operators to switch to minority client implementations. The protocol dynamics rapidly drives the network from an initial supermajority to a balanced, uniform distribution.
This convergence is stable over time, with node operators continuously responding to dynamic economic incentives, resulting in sustained client diversity.
Under the assumption of full rationality, the protocol thus eliminates supermajority risks and the catastrophic effects of worst-case scenarios.
Our verifiable client diversity protocol achieves the goal of improving systemic blockchain network resilience.
\end{tcolorbox}

\section{Discussion}

\subsection{Open Challenges}

The implementation of verifiable client diversity introduces several technical challenges that need to be addressed. For zkVM-based proofs, the primary challenge lies in identifying a suitable instruction subset that is both unique enough to serve as a reliable fingerprint and practical for execution within existing client implementations. The computational overhead of proof generation, particularly in terms of speed and resource requirements, presents another significant challenge. Additionally, there are inherent risks associated with proof reuse that must be mitigated through appropriate freshness criteria or nonce mechanisms.

The TEE-based approach faces its own set of challenges. Identifying program subsets that can be executed efficiently and securely within TEE constraints requires careful consideration of the hardware limitations. The performance overhead and latency associated with attestation generation must be minimized to maintain system efficiency. Furthermore, the reliance on centralized hardware manufacturers for attestations raises important questions about trust and decentralization that must be carefully evaluated.

\revised{In this paper, we model node operators as economically rational with respect to participation rewards, as related work \cite{pavloff2024game}.
In practice, validators face non-trivial switching costs and operational risk constraints.
They also consider other economically adjacent factors, such as the maturity of each client and compatibility with their stack.
These frictions can slow down immediate responses to incentives.
A more comprehensive model of validator behavior is left to future work.

Finally, we acknowledge a limitation of our evaluation: the client diversity simulation uses synthetically modified variants of a single client. While appropriate for isolating the economic and verification mechanisms in a controlled setting, this setup cannot fully capture cross-implementation nuances that arise among independently developed clients. However, our design and proofs-of-execution are implementation-agnostic: commitments are defined per implementation and per protocol step, and the verifier only checks cryptographic identities without assuming any specific codebase. As future work, we plan to extend the evaluation to multiple client implementations.
}

\subsection{Alternative Reward Protocols}

The design of the client diversity protocol offers flexibility in how rewards are distributed. One alternative approach is to integrate the reward mechanism directly within the blockchain protocol itself, rather than implementing it as a separate smart contract. This could be achieved through existing channels between the execution and consensus layers, such as Ethereum's execution-consensus interface, potentially reducing complexity and gas costs while maintaining the core benefits of the reward system. As shown in our evaluation of RQ2, the cost of verifying proofs and distributing rewards via a smart contract can be significant, especially for TEE-based attestations, which incur high gas fees. By integrating the reward mechanism directly into the blockchain protocol, these gas costs can be entirely bypassed, making the incentive mechanism more economically efficient and further improving the practicality and scalability of the client diversity protocol.

\subsection{Potential Drawbacks of Verifiable Client Diversity}

While verifiable client diversity offers significant benefits for enhancing blockchain resilience, it introduces several technical considerations that must be carefully managed. The proof generation process, particularly with zkVMs, can be computationally intensive and may introduce latency in client operations. This could potentially impact the overall performance of the network if not properly optimized. Additionally, the hardware requirements for TEEs may limit participation to nodes with compatible hardware, potentially affecting network decentralization.

Privacy considerations also play a crucial role in the design of verifiable client diversity. The cryptographic proofs generated by nodes may inadvertently leak information about validator identities, potentially exposing them to targeted attacks. Malicious actors could exploit knowledge of a validator's setup to launch denial-of-service attacks or target known vulnerabilities. These privacy risks must be carefully balanced against the benefits of verifiable execution.

\subsection{Governance and Financial Incentives}
\label{sec:governance}

The financial incentive mechanism for minority clients must be carefully designed to achieve the desired balance between encouraging diversity and maintaining network stability. The system should provide sufficient rewards to make running minority clients economically attractive while ensuring that the rewards remain sustainable in the long term. Additionally, the protocol must include mechanisms to discourage fraudulent submissions and malicious behavior, such as penalties for false claims or attempts to manipulate the client distribution.

The governance of these financial incentives requires careful consideration of several parameters, including the maximum and minimum reward amounts, the rate at which rewards adjust based on client distribution, and the criteria for determining which client implementations qualify for rewards. These parameters must be set through a transparent and inclusive governance process that involves all relevant stakeholders, including client developers, validators, and the broader blockchain community.

\noindent\textbf{Protocol Parameters.}
The primary governance decisions include determining the reward amounts for minority client implementations and establishing a fair distribution mechanism that encourages diversity while maintaining network stability. Additionally, defining appropriate penalties for fraudulent proof submissions or malicious behavior is crucial for maintaining protocol integrity.

\revised{Another critical governance decision involves determining the number of client implementations to support; and establishing criteria for including new implementations or removing outdated ones from the diversity protocol. This includes establishing processes for managing client version updates. This important point calls for dedicated software governance research.}

\noindent\textbf{Governance Process.}
The governance process should be transparent and inclusive, allowing for regular review and adjustment of protocol parameters based on network conditions and implementation distribution. Clear communication channels must be established for stakeholders to propose changes and participate in decision-making. Furthermore, mechanisms for emergency updates should be in place to address critical vulnerabilities or network issues.

This governance framework ensures the protocol remains adaptable to changing network conditions while maintaining its core objective of promoting client diversity.

\section{Related Work}
Blockchain client diversity finds its roots in the seminal work of Avizienis~\cite{avizienis1985n}, who introduces the concept of N-version programming to achieve fault tolerance in software systems.
It focuses on reliability through multiple independently developed software versions.
To our knowledge, our work is the first to bridge N-version programming with proos of execution and economic incentives.

Additionally, there are several works that leverage existing client diversity for better resilience in the context of blockchain.
For instance, Isenkul~\cite{proof-of-diversity} proposes a proof-of-diversity consensus protocol, where many diversity dimensions are taken into account for choosing the blockchain's validator set.
Tools such as Fluffy~\cite{yang2021finding} and Etherdiffer~\cite{kim2023etherdiffer} use differential testing of diverse Ethereum implementations to detect bugs.
N-ETH~\cite{neth} proposes an N-Version setup to increase availability of blockchain clients.
However, none of these works utilize a verifiable computing approach for proving code diversity as we do.

Related to measuring client implementation distribution,
the Nethermind's team conducts theoretical analysis on three methods to allow Ethereum validators to self-report their client diversity data~\cite{allow_clientinfo}.
Furthermore, Blockprint~\cite{blockprint} adopts a machine learning strategy to identify types of clients on Ethereum's consensus layer.
The Chainbound team proposes a methodology to measure the geographical distribution of the validators~\cite{validator_decentral}.
Ryan et al.~\cite{ryan2025recommendation} propose a client recommendation algorithm for validators/operators based on programming languages, cryptographic libraries, databases, development teams, and client distribution.
However, these works do not provide any guarantees on the collected information. 
In contrast, we aim to create the first solution to identify the used client implementations in a provable way. From there, our client diversity protocol would balance client distribution by integrating it with economic incentives. 

In the realm of blockchains and verifiable computation,
Arbitrum Nitro~\cite{Arbitrum_Nitro} introduces a mechanism to prove correct execution of EVM code by using refereed delegation of computation.
Similarly, Specular~\cite{ye2024specular} proposes an optimistic Ethereum rollup that supports interactive fraud proofs for diverse client implementations.
TrueBit~\cite{teutsch2024scalable} is a scalable verification solution for blockchains that employs a combination of economic incentives and a dispute resolution mechanism.
It enables secure outsourced computation while addressing the verifier's dilemma, thus supporting larger-scale task processing~\cite{teutsch2024scalable}.
Sepideh et al.~\cite{avizheh2024refereed} proposes a universal composability framework for verifiable computation.
They adopt smart contacts to ensure integrity when computation is outsourced to multiple servers.
Castillo et al.~\cite{castillo2025trustedcomputeunitsframework} present a system for trustless cross-organization pipelines using both distributed ledgers and verifiable computation for integrity and confidentiality.
While these works employ verifiable computation for their specific scenarios, none of them focuses on reliability of blockchain infrastructure. 

\section{Conclusion}

This paper introduces a framework to address the skewness of client implementations in blockchains by leveraging financial incentives and interactive fraud proofs. 
Our approach aims to create a more balanced distribution of clients, enhancing the network's resilience and decentralization. 
We implement the framework and empirically evaluate its effectiveness, providing insights into its impact on blockchain security and stability. 
This work offers a promising step toward mitigating monoculture risks and fundamentally improving the reliability of decentralized systems.

\section*{Acknowledgments}
This research has been supported by the Ethereum Foundation, the CHAINS project funded Swedish Foundation for Strategic Research (SSF), and by the Wallenberg Autonomous Systems and Software Program (WASP) funded by the Knut and Alice Wallenberg Foundation.

\bibliographystyle{unsrt}
\bibliography{references.bib}

\end{document}